\documentstyle[prd,aps,floats,psfig,preprint]{revtex}
\begin{document}
\title{ Discriminating signal from background  using neural networks. 
Application to  top--quark search at the Fermilab Tevatron
\thanks{This research is partly
supported by EU under contract number CHRX-CT92-0004 and by the
Comissionat per Universitats i Recerca de la Generalitat de Catalunya.}
}
\author{Ll. Ametller$^a$, Ll. Garrido$^{b,c}$, G.
Stimpfl--Abele$^{a,b}$, P. Talavera$^a$ and P. Yepes$^d$ }
\address{$^a$Departament de F\'\i sica i Enginyeria Nuclear, 
Universitat Polit\`ecnica de Catalunya, E-08034 Barcelona, Spain}
\address{$^b$Departament Estructura i Constituents Mat\`eria, Universitat de
Barcelona, E-08028 Barcelona, Spain}
\address{$^c$Institut de F{\'\i}sica d'Altes Energies, Universitat Aut\`onoma
de Barcelona, E-08193 Bellaterra (Barcelona), Spain}
\address{$^d$Rice University, Houston, TX 77251-1892, USA}
\date{\today }
\maketitle
\begin{abstract}
The application of Neural Networks in High Energy Physics to
the separation of signal from background events is
studied. A variety of problems usually encountered in this
sort of analyses, from variable selection to systematic
errors, are presented.  The top--quark
search is used as an example to illustrate the problems and proposed
solutions.
\end{abstract}


It is well known that neural networks (NN's) are useful tools for pattern
recognition. In High Energy Physics, they have been
used or proposed as good
candidates for tasks of signal versus background classification. 
However, most of the existing studies 
are somewhat academic, in the sense that they essentially compare the
NN performances with other classical techniques of classification
using Monte Carlo (MC) events for that purpose. In realistic applications,
real events should be analyzed and compared with simulated events,
introducing systematic effects
which have to be taken into account and 
could significantly modify the efficiency of the analysis. 
We try to give some insight in this
direction using the top quark search at the Fermilab Tevatron as 
illustration.  
The top quark has been observed by the CDF \cite{CDF}
and D0 \cite{D0} collaborations. 
Recently, NN's have been applied 
to experimental top quark searches by the D0 Collaboration \cite{Push},
for a fixed top quark mass,
concluding that 
NN's are more efficient than traditional methods, 
in agreement with previous parton level studies \cite{AGT}.

In this paper we continue and complete the analysis of Ref. \cite{AGT} for 
the top quark search at the Tevatron.
A more realistic study is performed by including parton hadronization
and detector simulation with jet reconstruction.
In addition, contrary to
Ref.\cite{AGT} where the top mass was fixed, the present study is valid for
a large range of top mass values.
Moreover, the number of kinematical
variables considered is enlarged
and different ways of selecting subsets of the most
relevant ones to the process under consideration are discussed.
Finally, the influence of
systematic errors on the NN results is studied.
 
 
The analysis is focused on the top quark search at the $p \bar p$
Fermilab Tevatron operating at $\sqrt{s}=1.8$ TeV.
The one-charged-lepton channel,
$ p \bar p\to t \bar t \to l \nu j j j j$ with $l= e^{\pm}, \mu^{\pm}$,
is considered as the signal to
look for.
 The main background is
  $ p \bar p \to W j j j j \to l \nu j j j j$.
Exact tree-level amplitudes with
spin correlations were used to generate MC samples for
both signal and background.
The latter was evaluated with VECBOS \cite{Vecbos}.
The CTEQ structure functions \cite{STRF} at the scale
$Q=m_t$ ($Q= <p_t>$) for the top signal (background) were utilized.
The LUND fragmentation model \cite{lund} was used to hadronize the quarks
and/or gluons. The obtained events were passed through
a fast MC program which simulates the
segmentation of a D0--like calorimeter.
Jets are reconstructed with a simple algorithm
based on the routine
used in the LUND package and electrons are defined as isolated
clusters with more than 90\%
electromagnetic energy.

Uncorrelated MC signal samples were generated for top masses
$m_t=150$,$168$, $174$, $189$ and $200$ GeV. 
Events with one-charged-lepton and four jets
satisfying  the  following acceptance cuts were selected:
$ p_t^j, p_t^l, p\llap /_t > 20   \  \hbox{\rm GeV}; 
|\eta^j|, |\eta^l| < 2  $ and 
$\Delta R_{jl}, \Delta R_{j j} > 0.7$. The symbol
$p_t$ ($\eta$) stands for transverse momentum (pseudorapidity)
and the indices $j=1,4$ and $l$ refer to the four jets and charged
lepton respectively; $p\llap /_t$ is the missing transverse momentum
associated with the undetected neutrino and
$\Delta R=\sqrt{(\Delta \eta)^2+(\Delta \phi)^2}$ is the
distance in the $\eta-\phi$ space, where $\phi$ is the azimuthal
angle.
The cross sections after the acceptance cuts for the signal and
the  background are given in Table \ref{table1}.

\begin{table}
\begin{center}
\begin{tabular}{|c|c|c|c|c|c|c|}
 \hline
$m_t \hbox{\rm (GeV)}$ & $150$ & $168$ & $174$ & $189$ & $200$ & backg \\
\hline
$\sigma $(pb) & 0.63 & 0.39 & 0.31 & 0.21 & 0.16 & 0.89 \\
 \hline
\end{tabular}
\end{center}
\caption{Signal and background cross sections after the
acceptance cuts.}
\label{table1}
\end{table}


In order to use NN's as signal/background classifiers, we considered
layered feed--forward NN's with topologies $N_i \times N_h \times N_o$,
($N_i$, $N_h$ and $N_o$ are the number of input,
hidden and output neurons, respectively), with 
back--propagation as the learning algorithm to minimize a quadratic 
output--error.
Using a set of physical variables as inputs and
taking the desired output as 1 for
signal events and 0 for background events, the network output gives, 
after learning, the conditional probability that new test events
are of signal or background type \cite{LLUIS,Lluis1}, 
provided that the signal/background ratio used in the
learning phase corresponds to the real one.

The robustness of the NN method is shown by making
the results independent of the top mass, using several values
in the learning and testing phases.
During the learning phase a general network (GN)
is fed with a set of events which contains a signal
sample, composed by three subsamples corresponding to $m_t=150$,
$174$ and $200$ GeV, and a background sample
in a $1:1$ proportion. In so doing, the NN output 
loses its direct Bayesian 
interpretation when applied over data 
whose signal/background proportion is not $1:1$. 
Nevertheless,
the NN is still useful for classification \cite{LLUIS}.
This way of proceeding has been shown to
 optimize the learning process and allows to use the network in a
wide interval for the masses of the signal \cite{Georg}.
 
A set of $N=15$ initial variables was considered.
Some of them are chosen specifically to pin down the
{\it a priori} main characteristics of the top signal, while others are not
specific to the signal. For each reconstructed event we compute:
(1)  $S$, the sphericity;
(2)  $A$, the aplanarity;
(3)  $m_{W_{jj}}$, the invariant mass of the hadronically decaying $W$;
(4)  $p_t^{W_l}$, the transverse momentum of the leptonically decaying
 $W$;
(5)  $E_T$, the total transverse energy;
(6)  $p_t^l$, the charged lepton transverse momentum;
(7)  $\eta_l$, the charged lepton pseudorapidity; 
(8-11)  $p_t^i$, $i=1,4$, the transverse momenta of the jets in
 decreasing order and
(12-15)  $\eta_i$, $i=1,4$, the jet pseudorapidities in decreasing order.
The missing transverse-momentum has been assigned to the undetectable
neutrino and 
its longitudinal momentum inferred 
along the lines suggested in Ref.\cite{BOP2}

\begin{figure}[t]
\centerline{\psfig{figure=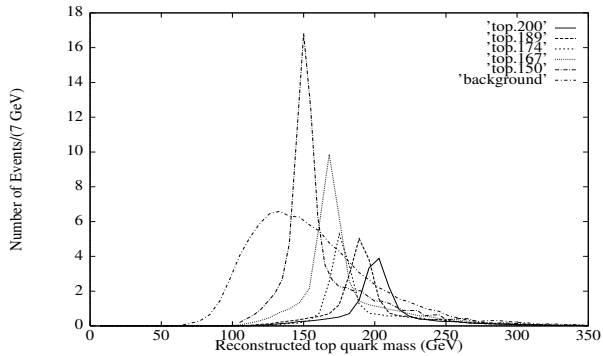,height=5cm,width=8cm,angle=-90}}
\caption{ Reconstructed top mass distribution
for several top signals and the background for 
${\cal L}= 100~\hbox{\rm pb}^{-1}$.}
\label{fig2}
\end{figure}

In the testing phase, the GN with topology $15\times 15\times 1$ is fed with
new background and top events. The latter can be chosen with masses
either corresponding to the values used
for learning or to new values $m_t=167$ or  $189$ GeV.
This differs from previous works
\cite{howard,AGT} where the same mass values were used in both 
learning and testing steps.
Figure \ref{fig2} shows the reconstructed top mass obtained for five top
signals and the background, corresponding to an integrated
luminosity  ${\cal L}=100 ~\hbox{\rm pb}^{-1}$. A good top reconstruction
is achieved for all masses considered 
but there is a substantial background contribution.
To further appreciate the GN's usefulness, 
five specialized NN's (SN) were trained with a top mass specific to
each one of them  and a generic background  common to all NN's.
Again, a $1:1$ signal to background ratio was used for learning.
The GN  and SN average errors, shown in Table \ref{table2}, 
are similar for all masses considered. 
This indicates that the GN performs fairly
well for a wide range of top mass values and, in
particular, for those never used in the learning phase.
Nevertheless, it is clear that the window for the
top mass should be reduced if the mass is more precisely known.
 
\begin{table}
\begin{center}
\begin{tabular}{|c|c|c|}
\hline
$                    $ & General Net & Specialized Net \\  \hline
$m_t \hbox{\rm (GeV)}$ & (GN)        &  (SN)           \\  \hline
 150 * & 0.12 & 0.10  \\  \hline
 167 ~ & 0.12 & 0.10  \\  \hline
 174 * & 0.11 & 0.10  \\  \hline
 189 ~ & 0.11 & 0.09  \\  \hline
 200 * & 0.10 & 0.07  \\  \hline
\end{tabular}
\end{center}
\caption{Average error per event. The asterisks indicate the top mass
values used in the General Network training.
}
\label{table2}
\end{table}

As a complementary check to the  present analysis, we have passed
the first top candidates ---published by CDF \cite{CDF1}---
through our initial $15\times 15 \times 1$ network in order 
to see wether they are compatible with our simulated signal
and/or background. 
Although our NN was trained with the simulation of the D0 detector,
such a check is still valid, since CDF quotes the parton level
momenta assigned to their top candidates.
One can therefore process those events through
our D0 detector simulation, reconstruct the variables used
in our analysis and obtain the individual output for the published
CDF top candidates.
The results are shown in Table \ref{table6}. It can be seen that
most of them give values close to 1,
showing that they are more compatible with our signal simulation
than our simulated background.

 
\begin{table}
 \begin{center}
\begin{tabular}{|c|c|}
\hline
Event number/Run & Net Output\\  \hline
 44414/40758   & 0.98  \\  \hline
 47223/43096   & 0.82  \\  \hline
 266423/43351   & 0.66  \\  \hline
 139604/45610   & 0.90  \\  \hline
 54765/45705   &  0.92 \\  \hline
 123158/45879   &  0.76 \\  \hline
 31838/45880   &  0.58 \\  \hline
\end{tabular}
 \end{center}
\caption{NN output for published CDF events.}
\label{table6}
\end{table}
 
 
The selection of the most relevant variables for a given process is one of
the major problems in experimental analyses.
Too many variables may introduce noise and make the event selection
task very difficult.
On the other hand, too much sensitivity may be lost when too few variables
are used.
In general, a large number of variables, $N$, can be considered and measured
for an event.
All $N$ variables carry some information on signal
versus background differences, but it is obvious that some subset of them
will be more valuable than other subsets for the separation task.
Therefore the selection of a subset with the `best' variables $n$ ($n < N$),
carrying the largest discrimination power between signal and background
samples, even if lower classification efficiencies may follow, is of interest.

In the  process of reducing the number of variables,
it is convenient to control the efficiency loss in
the classification task.
We suggest that NN's can be used for both the variable selection and the
evaluation of the
efficiency loss. For the former, there are several methods suggested in
the literature, some of which have been considered in the present analysis.
The latter will naturally be estimated in terms of the error function.
When reducing the number of variables, it is convenient 
to eliminate only a few variables in one step rather than making
multivariable rejection at once.
This introduces a mild dependence of the chosen variables
on the number of rejection steps, but turns out to be more efficient.
The following approach was adopted:
 
\begin{itemize}
\item{Step 1:} An $N\times N\times 1$ network is trained with the initial $N=15$
variables and its final error is computed, $E_{N}\equiv E_0$.
\item{Step 2:} A particular variable selection method is applied,
rejecting $n$ (keeping $N-n$) variables.
(It is convenient to choose small values for $n$.) 
\item{Step 3:} A new $(N-n)\times (N-n)\times 1$ network is trained
with the $N-n$
variables kept and its final error computed, $E_{N-n}$. If
the quantity $E_0/E_{N-n}$ is larger than, for instance, $75\%$, step 2
is repeated (replacing $N$ by $N-n$) to
further reduce the set of relevant variables. The algorithm stops if
$E_0/E_{N-n}< 0.75$.
This cut is arbitrary and the number of
selected variables depends on it. 
 
\end{itemize}

We have considered three methods involving weights for the selection
of the variables carried at step 2.
For every input neuron $k$, the following  quantities --in terms of
its connections with the hidden layer units, $w_{kl}$-- have been considered:
the sum of the weights \cite{LLUIS}, 
the variances \cite{Fogelman1} and the saliencies  \cite{Fogelman2}, defined 
respectively as
\begin{eqnarray}
\label{methods}
& {\hbox{Method 1:}} & \hskip 1cm
W_k=\sum_{l=1}^{N_h} |w_{kl}|,  \nonumber \\
& {\hbox{Method 2:}} & \hskip 1cm {\hbox{Var}}(k)= \frac{1}{N_h}
\sum_{l=1}^{N_h} w_{kl}^2
- \left(\frac{1}{N_h} \sum_{l=1}^{N_h} w_{kl} \right)^2 \nonumber
\\
& {\hbox{Method 3:}} & \hskip 1cm {\hbox{Sal}}(k)= \frac{1}{2}
 \sum_{l=1}^{N_h}
\frac{\partial^2 E}{\partial\omega_{kl}^2} \omega_{kl}^2
\end{eqnarray}
 

The surviving sets of relevant 
variables with error increase up to $25\%$ :
$3,5,8,10,11$ for methods 1 and 3, and 
$3,8,10,11,12,15$ for Method 2. The associated output-error
turns out to be $0.145$ and $0.178$ respectively.
At this stage, the set with the lowest
associated output-error, which corresponds to Methods 1 and 3, can be
safely chosen. The relevant variables are the mass of the hadronically 
decaying $W$, the total transverse energy $E_T$, and the jets transverse
momenta $p_t^1$, $p_t^3$ and $p_t^4$.
The quadratic error associated with 
this  set of five variables, obtained through systematic reduction, 
can be compared, for instance, with  the one obtained for the 
intuitive variables used in Ref.\cite{AGT}:
$S$, $A$, $m_{W_{jj}}$, $p_t^{W_l}$, $E_T$. The former is $18\% $ lower
than the latter, showing the usefulness of the methodical reduction.

We have trained an NN with the five relevant variables 
to study the enhancement of the
signal/background ratio as a function of the NN output cut.
For a specific cut, only
events with a network output higher than the specified cut are
selected.
Since the signal is peaked around 1 and the background around 0, 
it is clear that increasing the cut makes
the signal/background ratio larger.
A typical quantity that is used to reveal the existence of a signal
is the statistical significance, defined as:
$ S_s={N_s}/\sqrt{N_b} $,
where $N_s$ ($N_b$) is the number of signal
(background) events passing some NN output cut. 
It is assumed that $N_b$ can be estimated with
negligible error, but $N_s$ should be obtained from the actual
number of observed events, $N_o$, as $N_s=N_o-N_b$.
If both quantities $N_b$ and $N_s$ are large enough $(> 5)$,
$S_s$ can be interpreted as the number of standard deviations
that the background has to fluctuate  to obtain the observed number
of events. In such a case, the number of signal events
is also given by
$N_s=N_o-N_b\pm \sqrt{N_o}$.

\begin{figure}[t]
\centerline{\psfig{figure=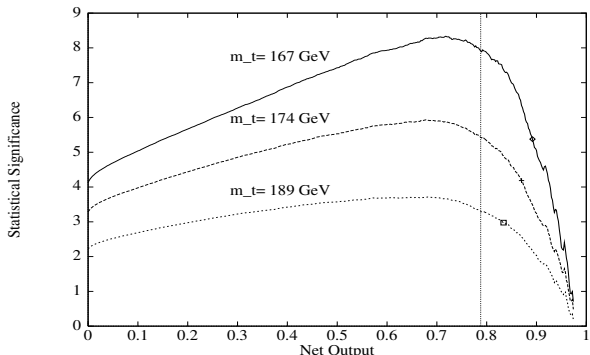,height=5cm,width=8cm,angle=-90}}
\caption{The Statistical significance as a function of the cut on
the NN output. The
symbols on the curves and the vertical line indicate the maximum network
output cuts such that more than five signal and five 
background events survive, respectively}
\label{fig3}
\end{figure}

\begin{figure}[t]
\centerline{\psfig{figure=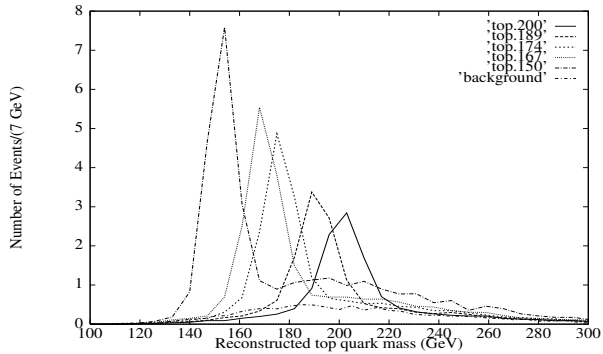,height=5cm,width=8cm,angle=-90}}
 \caption{Reconstructed top mass distribution
 for several top mass signals and  the background, for events with
 outputs larger than $0.7$ and ${\cal L}=100 ~\hbox{\rm pb}^{-1}$.}
 \label{fig5}
 
\end{figure}

Figure \ref{fig3} shows the $S_s$    
for $m_t=$ 168, 174 and 189 GeV and 
${\cal L}=100 ~\hbox{\rm pb}^{-1}$.
Conservative limits
of validity are shown in the figure.
The vertical line at network outputs $\simeq 0.8$ indicates the maximum network
 output
cut such that $N_b\ge 5$.
In a similar way, the symbols on the curves
indicate the maximum output cut such that more than five
signal events still survive.
NN output cuts
between $0.6$ and $0.8$ increase  the ratio
signal/background with a minimal loss on the
signal and a significant loss on the background.
 Figure \ref{fig5} shows the reconstructed top mass with only those
 events with the NN output larger than $0.7$.  As can be observed
 the signals dominate clearly over the background.

At this point, one can wonder about the benefits of using a reduced number of 
variables in the analysis. The main reason is to avoid 
possible noise when a large number of variables is used. 
In fact, the allowed increase of $25\%$ for the average error translates
into decreases for the signal efficiency and statistical significance.
We have found that the efficiency (statistical significance) 
diminishes from $0.75$ ($6.8$) to $0.58$ ($6.0$) when reducing
from the initial $15$ to the final $5$ variables, for an NN output cut
of $0.7$, value chosen because it maximizes the statistical significance.
These can be considered dramatic losses.
However, our initial number of variables, $N=15$, was moderate and we could
optimize the NN learning avoiding local minima. 
In general, this can be done for small sets of variables,
but it is very difficult for large ones, thus being possible
that  NN's trained with small subsets of relevant variables reach 
better efficiencies and/or statistical significances
than NN's trained with larger variable sets.
 

We consider now  
some sources of systematic errors coming from eventual disagreements
between MC and real data. 
In standard analyses, where single cuts are
applied on single variables, the effects of systematic errors should be studied
only in the region around the cuts in an easy and well
understood way. In the case of an NN
the only possibility to study the systematic error in the classification is
to propagate the ``estimated'' systematic
errors on  the input variables to the
output. Two basic effects can be considered: shifts between data and
MC and different resolutions for the used variables.
We have studied the effect of  $2\%$
shifts and  $2\%$ change of resolution on the clusters energy. 
With these new energies
the five selected variables were reconstructed to obtain a ``new'' test
data to evaluate systematic effects.         
Notice that the $2\%$ variation of the reconstructed cluster energies 
has been chosen for illustration purposes. 
This procedure automatically 
includes the correlations of the NN input variables.
(There are  studies in the literature where this is not the case \cite{L3}.) 
The results depend on the NN output cut. In the region of interest, 
we have found that the
uncertainty due to systematic errors is comparable with  the uncertainty
coming from an error on  $m_t$ of $\pm \ 11$ GeV.

 
The application of Neural Networks to discriminate
signal from background in High Energy Physics has been studied,
using the top quark search at Fermilab as an example.
The 
analysis is valid for a large range of top mass values.
Special attention was paid to the
selection of  the most relevant variables.
Several methods --in terms of the
weights connecting the input and the hidden neurons--
were considered.
We conclude that Methods 1 and 3, making use of the sum of the weights
(in absolute value) and the weight saliencies, respectively, give similar
results and are more suited for the variable selection than Method 2,
using the weight variances.
The performance of the reduced  NN
was studied in terms of the statistical significance. 
When comparing it with the
initial NN, we found a small decrease for the
statistical significance, and moderate loss of the signal efficiency.
Finally, the effect of propagating systematic errors arising from
energy shifts and changes in resolution have been studied.
This automatically accounts for the correct correlations among
the inputs.

\end{document}